# Influence of cation vacancy concentrations on ultra-low thermal conductivity in (1-$x$)BiVO$_4$ – $x$Bi$_{2/3}$MoO$_4$ scheelite solid solutions

Guillaume F. Nataf[±], Hicham Ait Laasri, Damien Brault, Tatiana Chartier, Chalit Ya, Fabian Delorme, Isabelle Monot-Laffez, Fabien Giovannelli[*]

GREMAN UMR7347, CNRS, University of Tours, INSA Centre Val de Loire, 37000 Tours, France

[*] fabien.giovannelli@univ-tours.fr

[±] guillaume.nataf@univ-tours.fr

**Abstract**

Bismuth vanadate – bismuth molybdate solid-solution was prepared to elaborate ceramics with different amounts of cation vacancies. Dense ceramics with similar microstructures were obtained and the evolution of their melting point, specific heat, thermal diffusivity, and conductivity as a function of the amount of vacancy was evaluated. At room temperature, the thermal conductivity decreases from 1.74 W m$^{-1}$ K$^{-1}$ for BiVO$_4$ ($x$=0) to 1.12 W m$^{-1}$ K$^{-1}$ for Bi$_{0.867}$□$_{0.133}$Mo$_{0.4}$V$_{0.6}$O$_4$ ($x$=0.4). Moreover, we show that a very small amount of vacancy (1.7%, $x$=0.05) is enough to provide a large decrease in thermal conductivity by more than 15%, in agreement with a mass fluctuation scattering model. However, the temperature of the melting point also decreases with increasing amount of vacancy. Our results suggest adding only a very small amount of vacancy as the best strategy to obtain superior materials for thermal barriers and thermoelectric devices, with ultra-low thermal conductivity and high-temperature stability.

Simple oxides with low thermal conductivity are crucial elements for the development of thermal barrier coatings and thermoelectric devices [1–5]. Their thermal conductivity is limited by scattering of phonons with phonons, defects and boundaries [6]. In these systems, strategies to reduce the thermal conductivity rely often on increasing anharmonicity [1,7–9], nanostructuring [10–14] or introducing defects [2,15–17]. In particular, vacancies lead to strong phonon scattering through their absence of mass and the mismatch in volume they impose on the lattice [2,17]. Among them, oxygen vacancies, i.e. anion vacancies, are the most often used. In $Zr_{0.92}Y_{0.08}O_{1.96}$, state-of-the-art thermal barrier coating material, 2% of oxygen vacancies reduce the conductivity by 30% [2]. In aluminates, where 16.7% of the oxygen atoms are missing, the conductivity is even close to the low value of 1 W m$^{-1}$ K$^{-1}$ [18].

In contrast, cation vacancies have been less often used to reduce the thermal conductivity because they are less frequent in simple oxides. Still, in perovskites, they lead to large reductions in thermal conductivity [19–24]. For instance, the introduction of cation vacancies on the A-site of the thermoelectric La-doped $SrTiO_3$ ($ABO_3$ perovskite structure) decreases the conductivity at room temperature from 10 W m$^{-1}$ K$^{-1}$ to ~2 W m$^{-1}$ K$^{-1}$ for 27% of vacancies [19]. The overall temperature behavior also changes and becomes almost temperature independent with the addition of vacancies, similar to a glass-like behavior [19,23].

Interestingly, cation vacancies have been proposed to also reduce the thermal conductivity in scheelites that already exhibit ultra-low thermal conductivities [25–27]. This low thermal-conductivity, observed experimentally in several compounds [25,28–30], arises from the intrinsic structure of $ABO_4$ scheelites [27], consisting in the stacking of weak units [$AO_8$] and strong units [$BO_4$], and is reduced further in compounds with vacancies. As such, to date, the lowest conductivity in a scheelite, with values around 0.7 W m$^{-1}$ K$^{-1}$ has been obtained for $La_{0.67}\square_{0.33}MoO_4$, which contains a significant amount of cation vacancy (33%) on the A-site [25].

Here, we investigate the thermal conductivity of the solid-solution (1-$x$)$BiVO_4$ – $x$$Bi_{2/3}MoO_4$, also written $Bi_{1-x/3}\square_{x/3}Mo_xV_{1-x}O_4$, where depending on the composition [31] the amount of cation vacancy on the A-site (written $\square$) can be tuned, such that the influence of different amounts of vacancies can be examined, including values as low as 1.7%.

$Bi_{1-x/3}\square_{x/3}Mo_xV_{1-x}O_4$ ceramics were prepared by conventional solid-state-reaction process from commercial precursors of $Bi_2O_3$ (99.9%), $MoO_3$ (99.9%) and $V_2O_5$ (99.9%) from ChemPur. Four compositions were investigated: $x=0$, $x=0.05$, $x=0.2$ and $x=0.4$. The obtained mixtures were grounded for 1h at 300 rpm in a tungsten carbide planetary ball mill (Retsch PM 100). They were then calcined in an alumina crucible, in a muffle furnace, at 650°C for 4h at a



heating rate of 5°C min$^{-1}$, and pressed into cylinders (10 mm in diameter, ~1.5 mm in height) in a steel die under a uniaxial pressure of 130 MPa. Conventional sintering in air was performed for 2h at different temperatures depending on the composition, at a heating rate of 5°C min$^{-1}$: 830°C for $x$=0 and $x$=0.05, 760°C for $x$=0.2 and 680°C for $x$=0.4. Sintering temperatures decrease with increasing amount of molybdenum in the composition, following the decrease in temperature of the melting point of the solid solutions. The composition corresponding to $x$=1 was not investigated as the ceramic sintered at 620°C exhibited a relative density below 85%, in agreement with previous observations [31]. Higher sintering temperatures for $x$=1 led to the melting of the ceramics. The density of pellets was measured at room temperature by the Archimedes method using a MS204TS/00 analytical balance (Mettler Toledo) in distilled water.

Phase determination was performed by X-ray diffraction measurements using a Cu-Kα radiation (D8 Advance Bruker X-ray diffractometer). X-ray diffraction patterns were collected at room temperature from 10° to 80° (2θ) with a step size of 0.02° and an acquisition time of 1 second per step. The obtained patterns were refined with the Rietveld method using the software Profex. The microstructure of as-sintered surfaces was investigated by scanning electron microscopy (Tescan Mira 3 SEM) operating in secondary electrons (SE) mode with an acceleration voltage of 20 kV. Before the measurements, all samples were gold-coated to avoid the accumulation of electron charges at the surface under the beam.

Samples were then mechanically polished with a Buehler EcoMet 30 polishing machine, under a force of 5 N. Silicon carbide abrasive disks and samples were rotating in the same direction at 150 rpm and 60 rpm, respectively. The final polishing step was performed with a Buehler 40-7220 microCloth and a 1 µm alcohol-based diamond suspension (Presi). The polished surfaces were observed with an optical microscope (Olympus BX60) operating in reflexion with a polarizer and an analyser, and an objective x100. The elementary composition was analysed by energy dispersive X-ray spectroscopy (EDX), on 1.5mm x 1.5mm area, using the scanning electron microscopy.

Differential thermal analysis (DTA) measurements were recorded on a Diamond TG/DTA (Perkin-Elmer), using platinum crucibles and ~50 mg of powder from crushed ceramics, at a heating rate of 10°C min$^{-1}$ in air. Heat capacities ($C_p$) were measured on ~50 mg of powder from crushed ceramics by differential scanning calorimetry (Netzsch STA 449 F3 Jupiter) in platinum crucibles and in nitrogen atmosphere. They were obtained by comparison of heat flows measured at a heating rate of 20°C min$^{-1}$, between an empty pan, the powder investigated, and a sapphire. The thermal diffusivity was measured with the laser flash method (Netzsch LFA 457 MicroFlash) from 23°C to 350°C in 50°C steps, under primary vacuum (10$^{-}$



$^2$ mbar). Samples were coated with a thin layer of graphite to improve absorption of the neodymium glass laser pulse on the front sides and to increase emission in the infrared on the rear sides.

EDX measurements reveal that there were no losses during the synthesis since the expected cation compositions for the four studied ceramics are obtained within a 2% accuracy (table 1).

Figure 1a shows the X-ray diffraction patterns of the four samples. For $x=0$, the crystal structure corresponds to the monoclinic system [space group I2/b (15)], with $a=5.196$ Å, $b=5.096$ Å, $c=11.710$ Å, and $\gamma=90.366°$, as expected [31,32]. With increasing Mo content (i.e. increasing $x$), the structure progressively evolves towards a tetragonal system [space group I4$_1$/a (88)], in agreement with previous measurements [31–33]. This symmetry change leads to the gradual merging of diffractions peaks (101) and (011) (Fig. 1b), (200) and (020) (Fig. 1c), etc. It is finalized for $x=0.2$, where peaks are fully merged, corresponding to $a=b=5.165$ Å and $c=11.693$ Å, similar to previous reports where the boundary between monoclinic and tetragonal phases was found to lie at $x=0.1$ [31]. From the lattice parameters, theoretical densities are calculated and compared to densities measured by the Archimedes method, leading to excellent relative densities of 96 +/- 1% for all samples (table 1).

High relative densities of the ceramics are confirmed by scanning electron microscopy images shown in Fig. 2, where the porosity is clearly low. These images reveal grain sizes around 5.3 µm, 2.5 µm, 2.7 µm, and 3.0 µm for $x=0$, 0.05, 0.2 and 0.4, respectively (averaged over 50 grains). The grain size distributions range from 1.8 µm to 15.3 µm, 0.8 µm to 5.4 µm, 0.9 µm to 5.2 µm, 1.4 µm to 5.3 µm, for $x=0$, 0.05, 0.2 and 0.4, respectively. For comparison, optical images of polished surfaces are shown, and reveal similar grain size distributions (Fig. 2, table 1). These results indicate that molybdenum most likely inhibits grain growth, with amounts as low as $x=0.05$.

The change in composition strongly influences the melting behavior of the ceramics, as shown in Fig. 3. The initially sharp peak at 938 K in DTA for $x=0$ decreases to 927 K for $x=0.05$, and gets broader with increasing values of $x$. For $x=0.2$, it peaks around 897 K but extends down to ~800 K. For $x=0.4$, the temperature of melting spreads from ~860 K to ~700 K, with a profile typical of a peritectic melting.

Figure 4a shows the specific heat of the four samples. They are identical within the 5% accuracy of standard heat measurements, and in agreement with first-principles calculations performed on monoclinic BiVO$_4$ [34]. They increase from 400 +/- 5 J K$^{-1}$ kg$^{-1}$ at 60°C to 460 +/- 15 J K$^{-1}$ kg$^{-1}$ at 350°C. For $x=0$, a decrease of the specific heat typical of a second-order



phase transition is observed at 250°C, corresponding to the monoclinic-tetragonal phase transition of BiVO$_4$ [35]. For $x$=0.05, this transition is lowered to ~120°C, in agreement with previous observations [31]. For higher Mo contents, the transition (if any) occurs at lower temperature [31], outside of our measurement range, as expected from the tetragonal symmetry obtained by X-ray diffraction at room temperature. The value of the specific heat at room temperature is extrapolated from our experimental data at 50°C, following the slope from first-principles calculations [34] (Fig. 4a – purple dot for all compositions, within the 5% accuracy of the measurements).

The thermal diffusivity of the four samples is shown in Fig. 4b. At room temperature, the thermal diffusivity is reduced from 0.68 mm$^2$ s$^{-1}$ for BiVO$_4$ ($x$=0), to 0.57 mm$^2$ s$^{-1}$ for Bi$_{0.983}$□$_{0.017}$Mo$_{0.05}$V$_{0.95}$O$_4$ ($x$=0.05), to 0.52 mm$^2$ s$^{-1}$ for Bi$_{0.933}$□$_{0.067}$Mo$_{0.2}$V$_{0.8}$O$_4$ ($x$=0.2), to 0.46 mm$^2$ s$^{-1}$ for Bi$_{0.867}$□$_{0.133}$Mo$_{0.4}$V$_{0.6}$O$_4$ ($x$=0.4). With increasing temperature, the thermal diffusivity decreases for all samples, due to Umklapp scattering that becomes the dominant mechanism at high temperatures [6], and the difference between the four samples is reduced. Still, at 350°C, the thermal diffusivity for $x$=0.4 is ~15% lower than the thermal diffusivity for $x$=0.

Based on the obtained experimental data for the specific heat ($C_p$), density ($\rho$) and thermal diffusivity ($\alpha$), the thermal conductivity ($\kappa$) is calculated as:

$$\kappa(T) = C_p(T) \times \rho(T) \times \alpha(T) \qquad (1)$$

For densities, values at room temperature were used since they vary by less than 1% in the temperature interval of interest [36]. The thermal conductivity is shown in Fig. 5a. For all samples, it lies between 1 and 2 W m$^{-1}$ K$^{-1}$, indicating that BiVO$_4$, even without vacancies, is an excellent thermal insulator [2]. This is related to the anharmonicity in the lattice, associated with the presence of trivalent Bi whose lone pair of unbonded electrons results in a highly nonlinear repulsive force between atoms [34,37]. At room temperature, the conductivity is reduced from 1.74 W m$^{-1}$ K$^{-1}$ for BiVO$_4$, to 1.43 W m$^{-1}$ K$^{-1}$ for Bi$_{0.983}$□$_{0.017}$Mo$_{0.05}$V$_{0.95}$O$_4$, to 1.30 W m$^{-1}$ K$^{-1}$ for Bi$_{0.933}$□$_{0.067}$Mo$_{0.2}$V$_{0.8}$O$_4$, to 1.12 W m$^{-1}$ K$^{-1}$ for Bi$_{0.867}$□$_{0.133}$Mo$_{0.4}$V$_{0.6}$O$_4$ (table 1).

With increasing temperature, three different behaviors are revealed: a decrease of the thermal conductivity for $x$=0, an almost constant value for $x$=0.05, and a slight increase for $x$=0.2 and $x$=0.4. For $x$=0, the thermal conductivity variation is close to a $1/T$ dependency, which is indicative of Umklapp processes at high temperature [6]. For other compositions, the evolution of the thermal conductivity with increasing temperature is typical of a phonon-glass behavior, where it follows the evolution of the specific heat, instead of asymptotically



approaching the lowest thermal conductivity values from above [38,39]. It is often indicative of localized vibrations (called locons) or non-propagating vibrations (called diffusons) [40,41], and has been observed before in perovskite oxides with disorder on their A-site [42,43]. With increasing temperature, Umklapp scattering becomes more and more important [6], and the difference between the four samples is reduced, reaching the experimental resolution limits of our experiments for $x$=0.05 and $x$=0.2.

In Fig. 5b, thermal conductivity values are replotted as a function of the amount of vacancy, revealing the clear decrease of the conductivity with increasing amount of cation vacancy in the investigated temperature range. Several factors that often reduce the thermal conductivity can be excluded. All the samples have a similar relative density (96 +/- 1%). Grain sizes are different (average sizes from 2.7 to 4.8 µm), but they are similar for $x$=0.05, $x$=0.2 and $x$=0.4. The mean grain size is even slightly higher for $x$=0.4 than for $x$=0.2 (3.1 µm and 2.7 µm, respectively), and the thermal conductivity for $x$=0.4 is still lower than for $x$=0.2. This may be because grain sizes stay in the micrometer-range (and not the nanometer-range), which is much larger than the mean free path of the phonons in these materials at these temperatures, and small differences in grain sizes have thus a minor influence on the thermal conductivity. For $x > 0.2$, the symmetry of the system changes to tetragonal, but since there is not a strong change in the thermal conductivity between the monoclinic and tetragonal phases for $x$=0 and $x$=0.05, the influence of this symmetry change appears negligible (Fig. 5a). The observed decrease in thermal conductivity is therefore attributed to the influence of cation vacancies on the A-site and of Mo atoms on the B-site, and the dominating mechanism is likely to be the scattering by vacancies, as observed in other oxides [19–24]. In this case, it is A-site vacancies since the conventional sintering was performed in air, hindering the presence of a large amount of oxygen vacancies.

To confirm this assumption, we evaluate the influence of cation vacancies and Mo atoms on the conductivity, by considering a mass fluctuation scattering model [44,45]. We thus neglect the influence of strain, as often done [44–46], which is consistent with the fact that the unit cell volume $V_{cell}$ of Bi$_{1-x/3}$□$_{x/3}$Mo$_x$V$_{1-x}$O$_4$ varies little with increasing $x$ (table 1). The average atomic mass variance on the A-site of the ABO$_4$ scheelite is given by:

$$\overline{\Delta M_A^2} = \left(1 - \frac{x}{3}\right)(M_{Bi} - \overline{M_A})^2 + \frac{x}{3}(M_{Bi} + 2 <\overline{M}>)^2 \qquad (2)$$

with $M_{Bi}$ the molar mass of bismuth (208.98 g mol$^{-1}$), $\overline{M_A}$ the stoichiometry weighted average of A-site average mass $\overline{M_A} = \left(1 - \frac{x}{3}\right)M_{Bi}$, and $<\overline{M}>$ the average mass of the compound. The



second term of the equation follows the model of Klemens for mass-scattering induced by vacancies [46].

On the B-site, the average atomic mass variance is given by:

$$\overline{\Delta M_B^2} = x(M_{Mo} - \overline{M_B})^2 + (1-x)(M_V - \overline{M_B})^2 \qquad (3)$$

with $M_{Mo}$ (95.95 g mol$^{-1}$) and $M_V$ (50.94 g mol$^{-1}$) the molar masses of molybdenum and vanadium respectively, and $\overline{M_B}$ the stoichiometry weighted average of B-site average mass $\overline{M_B} = xM_{Mo} + (1-x)M_V$. For all compositions, we find that $\overline{\Delta M_A^2}$ is ~18 to ~28 times larger than $\overline{\Delta M_B^2}$ (table 1). The reduction in conductivity is increasing when the average mass variance of the compound $<\overline{\Delta M^2}>$ increases (equation 4). Since $<\overline{\Delta M^2}>$ is given by the stoichiometry weighted average of each site average mass, $<\overline{\Delta M^2}> = \left(\overline{\Delta M_A^2} + \overline{\Delta M_B^2}\right)/6$, the fact that $\overline{\Delta M_A^2}$ is much larger than $\overline{\Delta M_B^2}$ confirms that scattering by vacancies is the dominant mechanism leading to a reduction in thermal conductivity in Bi$_{1-x/3}\square_{x/3}$Mo$_x$V$_{1-x}$O$_4$.

We also assess quantitively the influence of vacancies on the thermal conductivity [44,45]:

$$\frac{\kappa}{\kappa_{BiVO_4}} = \frac{\tan^{-1} u}{u}, \qquad u^2 = \frac{(6\pi^5 V_0^2)^{1/3}}{2k_B v_s} \kappa_{BiVO_4} \frac{<\overline{\Delta M^2}>}{<\overline{M}>^2} \qquad (4)$$

with $\kappa_{BiVO_4}$ the thermal conductivity of the ceramic of BiVO$_4$ at room temperature, $u$ the disorder scaling parameter calculated with: $V_0$ the volume per atom (i.e. the volume of the unit cell obtained by X-ray diffraction divided by the number of atoms in the unit cell), $k_B$ the Boltzmann constant, $v_s$ the average speed of sound (taken as 2999.65 m s$^{-1}$ [34]), $<\overline{\Delta M^2}>$ the average mass variance and $<\overline{M}>^2$ the squared average atomic mass [$<\overline{M}> = (\overline{M_A} + \overline{M_B} + \overline{M_O})/6$]. The obtained values are shown in Fig. 5b. For $x$=0.05, they confirm the experimental result that a small amount of vacancy already leads to a large decrease in thermal conductivity since vacancy concentrations as low as 1.7% are enough to reduce the thermal conductivity by 18% at room temperature. For higher vacancy concentrations, the model still matches qualitatively but overestimates the decrease in thermal conductivity in the temperature range investigated compared to experimental values that give a reduction by 36% at room temperature (from $x$=0 to $x$=0.4). This could be related to a competing reduction of the anharmonicity in the lattice, due to a reduction of the number of Bi$^{3+}$ lone pairs when increasing the amount of vacancy.

In conclusion, dense ceramics of (1-$x$)BiVO$_4$ – $x$Bi$_{2/3}$MoO$_4$ with different amounts of cation vacancies on the A-site were obtained. Thermal characterizations reveal an ultra-low thermal conductivity, reduced significantly with increasing amount of vacancy. At room temperature, a large reduction of more than 15% was obtained for vacancy concentrations as low as 1.7%, indicating cation vacancies as a promising route to decrease thermal conductivity in scheelites and oxides. Up to 150°C, there is still a significant difference in thermal conductivity depending on the amount of cation vacancy. However, the temperature of the melting point decreases with increasing amount of vacancy. Our results suggest thus that only small amounts of vacancies should be used to obtain materials with ultra-low thermal conductivity while keeping high-temperature stability requirements.

**Conflicts of interest**

There are no conflicts to declare.

**Acknowledgments**

We thank Dr. Mael Guennou for discussions on $BiVO_4$. Funded by the European Union (ERC, DYNAMHEAT, N°101077402). Views and opinions expressed are however those of the authors only and do not necessarily reflect those of the European Union or the European Research Council. Neither the European Union nor the granting authority can be held responsible for them.

| $x$ (precursors) | 0 | 0.05 | 0.2 | 0.4 |
|---|---|---|---|---|
| Mo/(V+Mo) | 0 | 0.04 | 0.19 | 0.39 |
| Bi/(V+Mo) | 1.01 | 0.98 | 0.92 | 0.85 |
| Amount of vacancy (%) | 0 | 1.7 | 6.7 | 13.3 |
| $V_{cell}$ (Å$^3$) | 310 | 310 | 312 | 317 |
| Theoretical density (g cm$^{-3}$) | 6.94 | 6.91 | 6.79 | 6.59 |
| Experimental density (g cm$^{-3}$) | 6.68 | 6.53 | 6.51 | 6.39 |
| Relative density (%) | 96 | 95 | 96 | 97 |
| Mean grain size (μm) | 4.8 | 2.7 | 2.7 | 3.1 |
| Grain size distribution (μm) | 1.8 – 9.8 | 1.3 – 4.4 | 1.5 – 4.3 | 1.7 – 5.4 |
| $\kappa$ at room temperature (W m$^{-1}$ K$^{-1}$) | 1.74 | 1.43 | 1.30 | 1.12 |
| $\overline{\Delta M_A^2}$ (g$^2$ mol$^{-2}$) | 0 | 1682 | 6809 | 13791 |
| $\overline{\Delta M_B^2}$ (g$^2$ mol$^{-2}$) | 0 | 96 | 324 | 486 |
| $\kappa_{model}$ (W m$^{-1}$ K$^{-1}$) | 1.74 | 1.27 | 0.84 | 0.63 |

Table 1. Characteristics of the four sintered ceramics: $x$ according to precursors compositions, cationic ratios based on atomic percent values obtained by EDX, amount of vacancy considering the composition Bi$_{1-x/3}$□$_{x/3}$Mo$_x$V$_{1-x}$O$_4$, volume of the unit cell $V_{cell}$ and theoretical density calculated from X-ray diffraction, experimental density obtained by the Archimedes method and the corresponding relative density, mean grain size and grain size distribution measured on optical microscopy images, thermal conductivity $\kappa$ at room temperature, average atomic mass variance on the A-site ($\overline{\Delta M_A^2}$) and the B-site ($\overline{\Delta M_B^2}$) of the ABO$_4$ scheelite, and the resulting calculated thermal conductivity $\kappa_{model}$ according to the mass fluctuation scattering model.

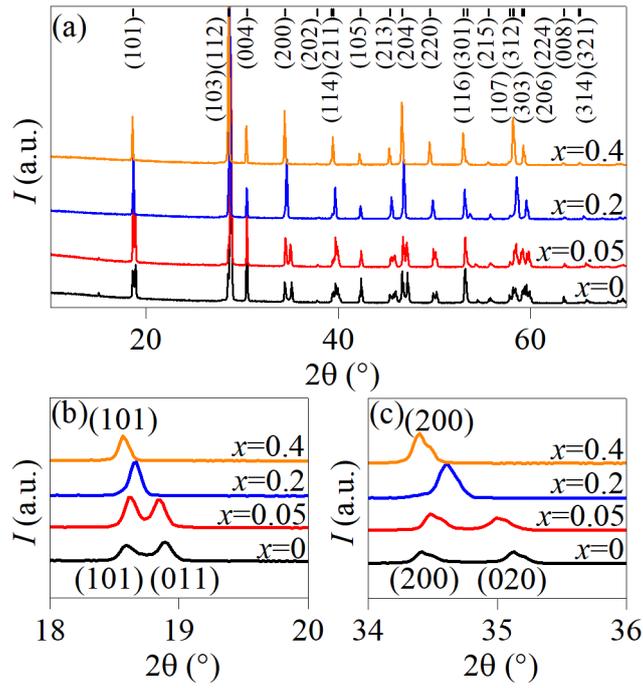

Figure 1. X-ray diffraction patterns obtained at room temperature for $Bi_{1-x/3}\square_{x/3}Mo_xV_{1-x}O_4$ with $x$ varying from 0 to 0.4, in the $2\theta$ range (a) 10° to 70°, (b) 18° to 20°, and (c) 34° to 36°. In panel (a) hkl indices of the reflections for the composition $x = 0.4$ are indicated.

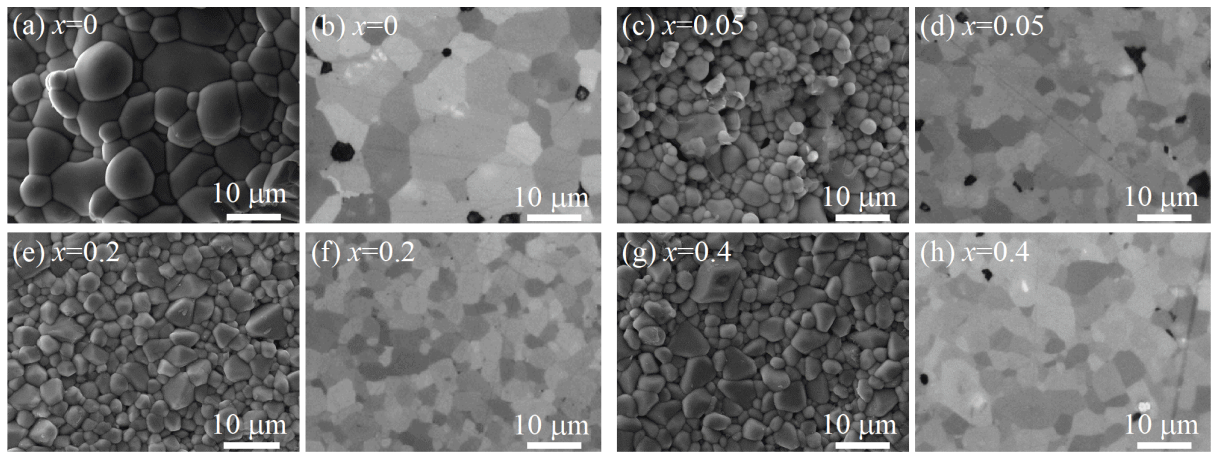

Figure 2. Scanning electron microscopy images of the surface of the sintered ceramics obtained with a secondary electron detector for (a) $x=0$, (c) $x=0.05$, (e) $x=0.2$ and (g) $x=0.4$. Optical images of the polished surfaces for (b) $x=0$, (d) $x=0.05$, (f) $x=0.2$ and (h) $x=0.4$

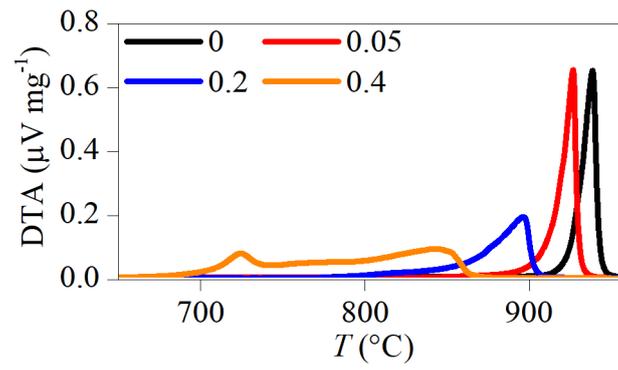

Figure 3. Differential thermal analysis as a function of temperature obtained on powder from crushed ceramics for *x*=0, 0.05, 0.2 and 0.4, after subtraction of the baseline.

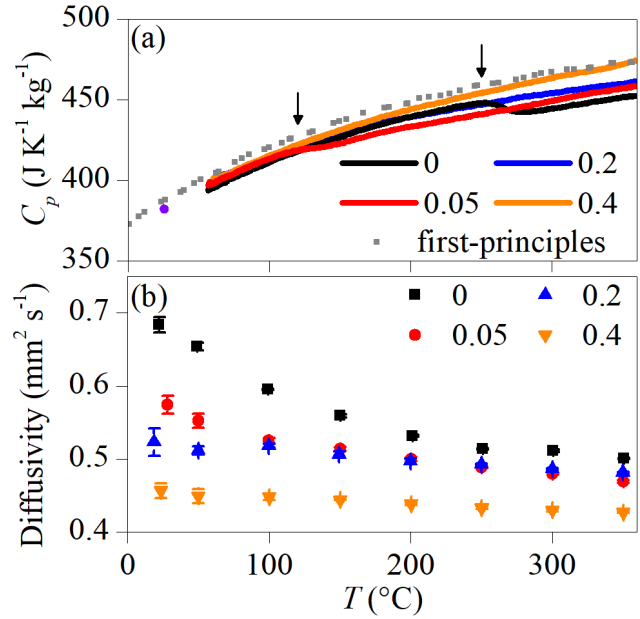

Figure 4. (a) Specific heat as a function of temperature obtained on powder from crushed ceramics for *x*=0, 0.05, 0.2 and 0.4 (for each sample, measurements were repeated three times and averaged). Grey dots indicate values obtained by first-principles calculations in ref. [34]. The purple dot at room temperature is extrapolated from our experimental data at 50°C, following the slope from first-principles calculations [34], and used for all compositions. Black arrows indicate decreases in specific heat capacities typical of second-order phase transitions. (b) Thermal diffusivity obtained on sintered ceramics for *x*=0, 0.05, 0.2 and 0.4. At each temperature, thermal diffusivity was recorded three times for averaging. At room temperature, an additional series of three diffusivity records was performed. Error bars indicate the standard deviation.

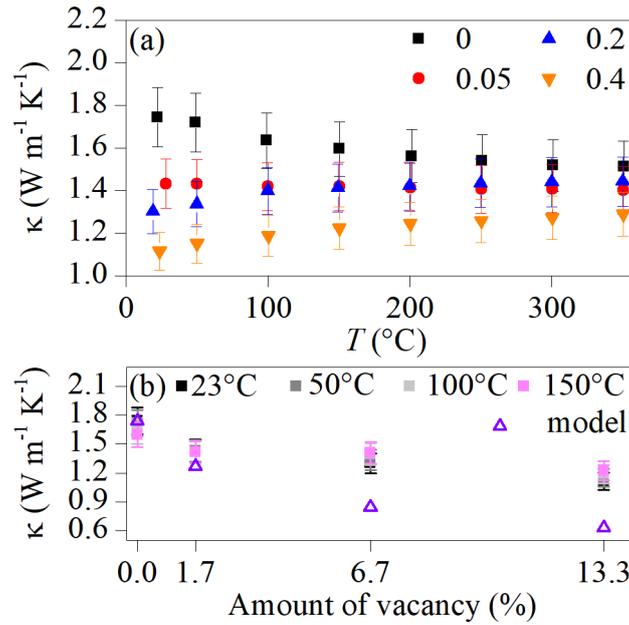

Figure 5. (a) Thermal conductivity as a function of temperature obtained on sintered ceramics for $x$=0, 0.05, 0.2 and 0.4. Error bars indicate an 8% error on the absolute value. (b) Thermal conductivity as a function of the amount of vacancy (equal to $x/3$) obtained on sintered ceramics at 23°C, 50°C, 100°C and 150°C, and values calculated using a mass fluctuation scattering model (empty purple triangles).